\documentclass[10pt]{emulateapj}
\usepackage{apjfonts}

\newcommand{\pfrac}[2]{\left(\frac{#1}{#2}\right)}
\newcommand{\gmr}{$\gamma$-ray~}
\newcommand{\gmrs}{$\gamma$-rays~}
\def\eps{\epsilon}
\def\d{{\rm d}}

\shorttitle{Prompt GeV emission from GRBs} \shortauthors{Li}

\begin{document}

\title{Prompt GeV Emission from Residual Collisions in Gamma-Ray Burst Outflows: Evidence from Fermi Observations of GRB 080916c}

\author{Zhuo Li\altaffilmark{1,2}}

\altaffiltext{1}{Department of Astronomy, Peking University, Beijing 100871, China}
\altaffiltext{2}{Kavli Institute for Astronomy and Astrophysics, Peking University,
Beijing 100871, China}

\begin{abstract}
The \gmrs from \gmr bursts (GRBs) are believed to be produced by
internal shocks driven by small timescale, $\sim1$~ms, variation
in the GRB outflows, and a pair-production spectral cutoff is
generally expected around the GeV range. However, the observed
optical flashes accompanying GRBs suggest that the delayed
residual collisions due to large timescale variation continue to
accelerate electrons. We show here that the inverse-Compton (IC)
scattering of the prompt \gmrs by these residual internal shock
electrons leads to a high energy emission beyond the previously
thought spectral cutoff, in agreement with the previous detections
of GeV photons by EGRET in several GRBs in conjunction with MeV
emission. We expect a spectral break due to the transition from
the primary to residual internal shock emission at the previously
thought spectral cutoff, and expect systematic time delays of high
energy photons relative to MeV emission, the discovery of which
would provide stringent constraint on the outflow properties, but
requires large enough collection of high energy photons by, e.g.,
Fermi and AGILE satellites.

The recent Fermi-detected bright GRB 080916c unambiguously shows
the shifting of the prompt emission toward later times as the
photon energy increases. The second-scale shifting at >100 MeV is
much longer than the MeV variability time, as predicted in the
residual collision model. The observations imply that there should
be emission above 70 GeV in the source frame, which may not be
produced by primary internal shocks but by IC emission in residual
collisions. With the method involving time delays of high energy
emission, the bulk Lorentz factor of GRB 080916c is determined to
be $\Gamma\sim300$.
\end{abstract}


\section{Introduction}
The prompt MeV \gmrs from a GRB are well believed to be produced
by a unsteady outflow which causes internal collisions between
different parts with various velocities, leading to kinetic energy
dissipation \citep{PX94,RM94}\citep[see, however,][]{LB03,NK08}.
The internal shock model can naturally explain both the
non-thermal spectra and the complicated light curves of GRBs. The
observed temporal variabilities of \gmr emission are believed to
be reflecting the activities of the central engine
\citep{SP97,Kobayashi97}. The internal shocks are expected to
generate/amplify magnetic field and accelerate electrons, leading
to MeV \gmrs by synchrotron emission \cite[see, e.g.][for a
review]{Waxman rev}.

GRBs are mainly observed in MeV range, the properties of high
energy, say, $>100$MeV, emission are not well understood, which,
however, might be very helpful in constraining the physics of the
GRB emission region. For example, the observed $>100$~MeV photons
in conjunction with MeV emission by EGRET in several GRBs,
suggesting that they can avoid the $\gamma\gamma$ absorption, have
leaded to the conclusion that the GRB outflow must be
relativistically expanding with a Lorentz factor of $\Gamma\ga100$
\citep[e.g.][]{Baring97}\citep[see also][]{Waxman rev}. As the
development in high energy \gmr observations, e.g. AGILE and Fermi
are being well operated, there are great interests on detecting
the high energy pair-production spectral cutoff in order to
constrain the size and Lorentz factor of GRB emission region (e.g.
\citet{Baring00,{Lithwick01}}; and recent detailed consideration
by \cite{Gupta07,Gupta08,Granot08,Murase08}).

However, a simple spectral cutoff may not exist.
\citet[][hereafter LW08]{LW08} had noticed that the frequently
observed prompt optical emission
\citep{Akerlof99,Blake05,Vestrand05,Vestrand06,Yost07} is above
the expected synchrotron-self absorption limit, and suggests a
relatively large size of optical emission region, compared to that
of MeV emission. Actually in the context of internal shock models
one would expect delayed collisions in the outflow following the
small radius collisions driven by smallest timescale variation of
the outflow properties, and these delayed collisions can naturally
account for the relatively bright optical emission (LW08). The
recently detected optical flash from the naked-eye GRB 080319b
\citep{Bloom08,DElia08,Racusin08,319bapj} appears to vary rapidly
in times and its temporal profile is correlated to the MeV
emission in second scales \citep{080319b_delay}, supporting that
the prompt optical emission in GRBs is produced by internal shocks
within the outflow, i.e. synchrotron emission from residual
collisions (Li \& Waxman, in preparation). The electrons in
residual collisions mainly cool by IC scattering off the MeV
photons, which produces high energy emission at large radii where
the optical depth due to pair production is reduced (LW08). We
consider here this high energy emission and show that it may
"smear" the previously thought pair-production spectral cutoff.
But a spectral turnover is still expected, which may be observed
by Fermi and AGILE although more difficult to detect than a simple
cutoff. The high energy emission from residual collisions is also
expected to be delayed relative to MeV \gmrs. It should be noticed
that we only focus on the {\em prompt} high energy emission which
appears simultaneously with the MeV \gmr emission.

We show first in \S 2 the strong $\gamma\gamma$ absorption during
the MeV \gmr emitting phase, next discuss in \S 3 the high energy
emission from residual collisions, then the model is applied to
the recent Fermi detection of GRB 080916c \citep{Fermi916c} in
\S4, which might have provided evidences of the model, finally a
general discussion on observations is given in \S 5.

\section{$\gamma\gamma$ absorption at small radii}
Consider a highly relativistic outflow with a bulk Lorentz factor
$\Gamma=10^{2.5}\Gamma_{2.5}$. The small timescale variation will
lead to strong internal shocks, which produce \gmr emission. Let
us denote the radius where \gmrs arise by $R_\gamma$. Due to
geometry effect, the observed fastest variability timescale
$t_{\rm var}\la10^{-2}$s \citep[e.g.][]{Woods95} in \gmr light
curves suggests that the size of \gmr emission region is limited
to $R_\gamma\la2\Gamma^2ct_{\rm var}$. If the \gmr emitting
electrons are fast cooling (with cooling time shorter than the
dynamical time of the outflow), we should take the equality, thus,
$R_\gamma\approx6\times10^{13}\Gamma_{2.5}^2t_{\rm var,-2}$cm,
where $t_{\rm var}=10^{-2}t_{\rm var,-2}$s. Actually, in the
context of internal shock models, we do not expect the magnetic
field is generated with energy density much higher than that of
accelerated electrons. In order to have synchrotron emission
peaking at $\eps_b\sim1$~MeV, as observed in GRBs, the radius of
the \gmr emitting region should be small (LW08),
$R_\gamma\la10^{13}L_{\rm bol,52}^{1/2}(\eps_b/1~\rm MeV)^{-1}$cm,
where $L_{\rm bol}=10^{52}L_{\rm bol,52}\rm erg~s^{-1}$ is the
bolometric \gmr luminosity.

Let us consider the $\gamma\gamma$ absorption due to pair production inside the GRB
source. For a photon of high energy $\varepsilon$ the optical depth to pair-production
within the GRB source is given by the product of the pair-production rate,
$1/t'_{\gamma\gamma}(\varepsilon)$, and the dynamical time, the time required for
significant expansion of the plasma, $t'_d\simeq R_\gamma/\Gamma c$ (primes denote
quantities measured in the outflow rest frame),
$\tau_{\gamma\gamma}(\varepsilon)\simeq R_\gamma/\Gamma
ct'_{\gamma\gamma}(\varepsilon)$. $t'_{\gamma\gamma}(\varepsilon)$ depends on the
energy density and on the spectrum of the radiation. The (outflow rest frame)
radiation energy density is approximately given by $U'_\gamma=L/4\pi
R_\gamma^2c\Gamma^2$. The GRB spectrum can typically be described as a broken power
law, $dn/d\eps\propto\eps^{-\beta}$, with $\beta\approx-1$ at low energy,
$\eps<\eps_b\sim1$~MeV, and $\beta\approx-2$ at $\eps>\eps_b$ \citep{Band93}. High
energy photons with energy $\varepsilon'$ exceeding
$\varepsilon'_b\equiv2(m_ec^2)^2/\eps'_b$, may produce pairs in interactions with
photons of energy exceeding $\eps'=2(m_ec^2)^2/\varepsilon'<\eps'_b$ (the rest frame
photon energy $\varepsilon'$ is related to the observed energy by
$\varepsilon=\Gamma\varepsilon'$). For $\eps'<\eps'_b$ we have
$dn/d\eps'\propto\eps^{\prime-1}$, which implies that the number density of photons
with energy exceeding $\eps'$ depends only weakly on energy. Thus, $t_{\gamma\gamma}$
is nearly independent of energy for $\varepsilon'>\varepsilon'_b$,
$t_{\gamma\gamma}^{\prime -1}\approx(\sigma_T/16)cU'_\gamma/\eps'_b$, which gives
\begin{equation}\label{eq:tau}
  \tau_{\gamma\gamma}(\varepsilon>\varepsilon_b)\simeq
    1.1\times10^2\frac{L_{52}}{t_{\rm var,-2}\Gamma_{2.5}^4} \pfrac{\epsilon_b}{\rm 1~MeV}^{-1}.
\end{equation}
Note, we have approximated the $\gamma\gamma$ cross section above
the pair-production threshold as $3\sigma_T/16$ \citep{Waxman
rev}. Also note that as the energy density $U_\gamma'$ (and hence
the related luminosity $L$) depends on the energy band considered,
the one used in calculating $t_{\gamma\gamma}^{\prime}$ is the
energy density below $2\times\eps_b'$. Hereafter, without special
emphasis the luminosity $L=10^{52}L_{52}\rm erg\,s^{-1}$ is the
so-called MeV luminosity, only corresponding to emission at
$<2\times\eps_b$, i.e., $L\equiv\int_0^{2\eps_b}L_{\eps}d\eps$.

Photons of lower energy, $\varepsilon<\varepsilon_b$, interact to produce pairs only
with photons of energy $\eps'>2(m_ec^2)^2/\varepsilon'>\eps'_b$. Since the number
density of these photons drops like $1/\eps'$, we have
$\tau_{\gamma\gamma}(\varepsilon<\varepsilon_b)\approx
(\varepsilon/\varepsilon_b)\tau_{\gamma\gamma}(\varepsilon>\varepsilon_b)$, i.e.
\begin{equation}
  \tau_{\gamma\gamma}(\varepsilon<\varepsilon_b)\simeq
    2.2\times10^{-3}\frac{L_{52}}{t_{\rm var,-2}\Gamma_{2.5}^6}\frac{\varepsilon}{1~\rm MeV}.
\end{equation}
The optical depth increases as photon energy increases. Photons with high enough
energy are absorbed in the emission region. A spectral cutoff is defined by
$\tau_{\gamma\gamma}=1$,\footnote{Because a low energy turnover at $\eps_a\sim1$~keV
is expected in GRB spectra due to synchrotron self absorption, very high energy
photons with $\varepsilon\ga10^{16}L_{52}t_{\rm
var,-2}^{-1}\Gamma_{2.5}^{-2}(\eps_a/1\rm ~keV)^{-1}$eV still can escape from the GRB
source (e.g. \cite{LW07}; see also \cite{razza04}).}
\begin{equation}\label{eq:cut1}
  \varepsilon_{\rm cut}^{(1)}\simeq0.46\frac{t_{\rm
  var,-2}\Gamma_{2.5}^6}{L_{52}} \rm ~GeV.
\end{equation}
A comment on the approximation of the $\gamma\gamma$ cross section
should be made here. Since both the cross section and the GRB
photon spectrum decrease rapidly with photon energy, the
approximation is excellent- for a GRB spectrum with
$\beta\approx-2$, it precisely produces the optical depth within
2\%, compared to a calculation with full cross section.

There is another restriction for the cutoff besides eq.
(\ref{eq:cut1}). In deriving the cutoff, eq. (\ref{eq:cut1}), the
optical depth is not self-consistently calculated since a GRB
spectrum extending to infinity without a high energy cutoff is
assumed. Given the two factors that GRB spectrum usually appears
to be a steep slope, with the photon number dominated by low
energy photons, and that the high ($\varepsilon'>2^{1/2}m_ec^2$)
and low ($\varepsilon'<2^{1/2}m_ec^2$) energy photons annihilate
each other one by one, we only expect high energy photons might be
totally attenuated by low energy ones, other than the other way
around. So the cutoff should be $\varepsilon_{\rm
cut}'>2^{1/2}m_ec^2$, which is not assured by $\varepsilon_{\rm
cut}^{(1)}$ in eq. (\ref{eq:cut1}) ($\varepsilon_{\rm
cut}^{(1)}/\Gamma<2^{1/2}m_ec^2$ might happen). We need to define,
in the GRB source frame,
\begin{equation}
  \varepsilon_{\rm cut}^{(2)}=2^{1/2}\Gamma m_ec^2\simeq0.23\Gamma_{2.5} \rm
  ~GeV.
\end{equation}
Note, this condition is also forgotten by many other authors who
calculate the cutoff energy assuming a no-cutoff GRB spectrum. The
GRB spectral cutoff is, therefore, expected at the maximum between
$\varepsilon_{\rm cut}^{(1)}$ and $\varepsilon_{\rm cut}^{(2)}$
\citep{LS04},
\begin{equation}\label{eq:cut}
  \varepsilon_0\equiv\varepsilon_{\rm cut}(R_\gamma)
  \simeq\max\left[0.46\frac{t_{\rm var,-2}\Gamma_{2.5}^6}{L_{52}},~0.23\Gamma_{2.5}\right]\rm
  GeV.
\end{equation}
A critical Lorentz factor where $\varepsilon_{\rm cut}^{(1)}=\varepsilon_{\rm
cut}^{(2)}$ is function of $R_\gamma$ (hence $t_{\rm var}$),
\begin{equation}
 \Gamma_c\simeq280(L_{52}/t_{\rm var,-2})^{1/5}.
\end{equation}
$\varepsilon_0$ sensitively depends on $\Gamma$ thus detection of
the spectral cutoff may be very useful in constraining $\Gamma$,
\begin{equation}\label{eq:Gamma_cut}
  \Gamma=\min\left[360\left(\frac{L_{52}}{t_{\rm var,-2}^{\rm ob}}\frac{\varepsilon_0^{\rm ob}}{\rm
  1~GeV}\right)^{1/6}(1+z)^{1/3},~
  1390\frac{\varepsilon_0^{\rm ob}}{\rm  1~GeV}(1+z)\right].
\end{equation}
Here the redshift factor has been included,  i.e. $t^{\rm
ob}=t(1+z)$ and $\varepsilon^{\rm ob}=\varepsilon/(1+z)$.

The extension of GRB spectra to $\ga100$~MeV and the
characteristic variability time, $t_{\rm var,-2}\la1$, have
implied $\Gamma_{2.5}\gtrsim1$, assuming the $\ga100$~MeV photons
are produced in the same time and place as the MeV photons. Since
thermal pressure acceleration in the initial fireball can not lead
to much larger Lorentz factors, and in internal shock model much
larger Lorentz factors would lead to synchrotron emission peaking
below MeV band (see discussion in the first paragraph of \S2),
$\Gamma_{2.5}\approx1$ is typically adopted \cite[e.g][]{Waxman
rev}. The exact values of $\Gamma$ would be determined by the
detection of the high energy cutoffs in GRB spectra by, e.g.
Fermi. We show below that the situation may be complicated by the
delayed large-radius emission from the outflow.

\section{Large-radius emission from residual collisions}
After the initial strong collisions at small radii driven by the
small timescale, $\sim1$~ms, variation in the outflow, there are
residual collisions continue to occur when the outflow is
expanding to large radii. As the velocity fluctuation in the flow
is being smoothed out by the on-going collisions, the delayed
collisions become weaker, and the postshock electron energy and
magnetic field are smaller, which give rise to synchrotron
emission at longer wavelengths. LW08 has well discussed the
dynamics of the residual collisions and naturally explained the
optical flashes from GRBs by this delayed residual emission. As
mentioned in LW08, the energy density in the delayed collisions is
dominated by the primary \gmr emission, so that the residual
emission is dominated by IC scattering of the prompt \gmr photons.
In what follows consider the IC emission.

\subsection{Dynamics}
Let us approximate the outflow by a sequence of $N\gg1$ individual shells. LW08 has
considered the simplified case: the shell masses are equal; the shells are initially
separated by a fixed distance $ct_{\rm var}$; the extent of the shells is much smaller
than the radius of the outflow; and the Lorentz factors of individual shells are drawn
from a random distribution with an average $\Gamma$ and initial variance
$\sigma_{\Gamma,0}^2<\Gamma^2$. The model may be more complicated, by adding more
degrees of freedom, however LW08 has shown that this simple case naturally accounts
for the observed properties of GRB optical flashes. Here we will consider this
simplified case. Moreover, we adopt the assumption that two shells merge into a bigger
shell after a collision, i.e. the full inelastic collision where the internal energy
generated is fully radiated. If the postshock electrons carry an energy close to
equipartition and the electrons cool fast, the internal energy is always radiated
significantly. In the case of a significant fraction of the internal energy being
dissipated in each collision, the dynamical evolution of the outflow has been proven
to be similar to the full inelastic case (LW08).

In the simple case under consideration, the variance of the velocities of individual
shells (in the outflow rest frame) evolves with the outflow radius $R$ as
$\sigma_v\propto R^{-1/3}$. The outflow energy that is associated with the fluctuation
of shell velocities (in the outflow rest frame) and hence may be dissipated decreases
as
\begin{equation}
  E_{\rm fluc}\propto\Gamma\sigma_v^2\propto R^{-2/3}.
\end{equation}

In general, it is naturally expected that there might be a wide
range of variability timescale, $\rm \sim1~ms-10~s$, in the flow
properties. Large timescale variabilities might lead to more
energy dissipated at large radii. Thus, the slope should be
flatter than $-2/3$. If a power-law description, $E_{\rm
fluc}\propto R^{-q}$, is still available, one may have $0<q<2/3$.
We carry a monte carlo simulation to demonstrate this point in the
appendix.

\subsection{High energy emission}
Based on the dynamical evolution, the emission from the residual collisions can be
further predicted. Taking the common assumptions that in internal shocks the postshock
electrons and magnetic fields carry fixed fractions, $\eps_e$ and $\eps_B$,
respectively, of the postshock internal energy, the characteristic Lorentz factor of
postshock electrons (in the outflow comoving frame) scales as
$\gamma_i\propto\eps_e\sigma_v^2\propto R^{-2/3}$, and the postshock magnetic field
scales as $B^2\propto\eps_B\sigma_v^2n_e\propto R^{-8/3}$ (the particle number density
scales as $n_e\propto R^{-2}$).

We demonstrate that the electrons in residual collisions lose most
of their energy by IC cooling. If the prompt \gmrs last a duration
$T$ (observer frame), the plasma is overlapped with these \gmrs
until the outflow expands to $R\ga2\Gamma^2cT\simeq(T/t_{\rm
var})R_\gamma$. LW08 showed that when the synchrotron emission
lies in the optical band, the radius is $R_{\rm
opt}\simeq10^2R_\gamma$. For typical observed values $t_{\rm
var}\la10^{-2}$s and $T\sim10$s, the optical radius is still
relatively small, $R_{\rm opt}<2\Gamma^2cT$. Therefore during the
phase of late residual collisions that we concern, the plasma is
immersed in the radiation bath of the prompt \gmrs. Both the
photon energy density $U_\gamma$ and the particle number density
$n_e$ drop as $\propto R^{-2}$ hence the ratio
$y=U_\gamma/(B^2/8\pi)\propto\sigma_v^{-2}\propto R^{2/3}$
increases with $R$. Because $y\sim1$ in the \gmr producing phase,
we have $y>1$ in residual collision phase, so the radiation energy
density dominates that of the magnetic field. Let us consider the
properties of IC emission.

\subsubsection{Spectrum}
Consider first the energy band into which energy is radiated. At
radius $R$, the prompt \gmr photons with typical energy $\eps_b$
are up-scattered by electrons with characteristic Lorentz factor
$\gamma_i$ to energy $\varepsilon_{\rm IC}\simeq
\lambda\gamma_i^2\eps_b\simeq
\lambda\epsilon_e^2(m_p/m_e)^2(R/R_\gamma)^{-4/3}\eps_b$. Here we
assume $\gamma_{i,0}\sim \epsilon_e(m_p/m_e)$ as the electron
Lorentz factor emitting MeV \gmrs, and $\lambda$ accounts for the
correction due to uncertain geometry effect. It will be shown in
appendix that the correction factor $\lambda$ is order unity even
in the case that the prompt MeV photons are strongly beamed in the
rest frame of the outflow at $R$. The characteristic scattered
photon energy is
\begin{equation}\label{eq:ICph_energy}
 \varepsilon_{\rm IC}^{\rm ob}\simeq9\lambda\epsilon_e^2\frac{\epsilon_b}{\rm 1~MeV}\pfrac{R}{10^2R_\gamma}^{-4/3}(1+z)^{-1}\rm GeV.
\end{equation}
The scattering might take place within slight Klein-Nishina
regime, $\gamma_i\epsilon_b'/m_ec^2\sim$ a few $>1$, at small
radii $R\sim R_\gamma$, where the energy of scattered photons is
instead $\varepsilon_{\rm
IC}=\Gamma\gamma_im_ec^2=\epsilon_e\Gamma(R/R_\gamma)^{-2/3}m_pc^2\simeq3\times10^2\epsilon_e(R/R_\gamma)^{-2/3}$GeV.

\begin{figure}
\includegraphics[width=\columnwidth]{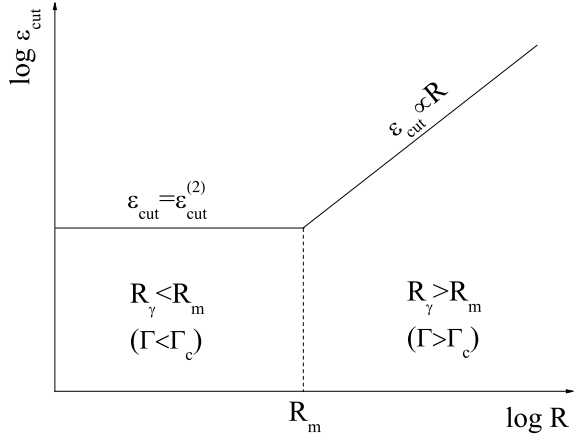}
\caption{Schematic plot of the attenuated energy evolving with
radius. There are two regimes. If the primary collisions that
produce MeV emission occur in the regime of $R_\gamma>R_m$ (i.e.
$\Gamma>\Gamma_c$), $\varepsilon_0=\varepsilon_{\rm cut}^{(1)}$
and $\varepsilon_{\rm cut}\propto R$ later. If $R_\gamma<R_m$
($\Gamma<\Gamma_c$), in the beginning the attenuated energy is a
constant, $\varepsilon_{\rm cut}=\varepsilon_{\rm cut}^{(2)}$ at
$R<R_m$, and turns to be $\varepsilon_{\rm cut}\propto R$ at
$R>R_m$.} \label{fig:cutvsR}
\end{figure}

Next consider the $\gamma\gamma$ absorption effect on the late
residual emission. For GRB outflow with $\Gamma>\Gamma_c$, the
initial cutoff energy for the primary emission is determined by
the first term in the bracket of eq. (\ref{eq:cut}),
$\varepsilon_0=\varepsilon_{\rm cut}^{(1)}$ . Eq. (\ref{eq:cut1})
implies that the spectral cutoff energy scales as
$\varepsilon_{\rm cut}\propto R\Gamma^4L^{-1}$, so the cutoff
energy increases with $R$ for fixed $L$. We have, for late
residual collisions, $\varepsilon_{\rm cut}\simeq\varepsilon_0
R/R_\gamma$. In the case of GRB outflow with lower Lorentz factor
$\Gamma<\Gamma_c$, the cutoff energy is initially a constant,
$\varepsilon_0=\varepsilon_{\rm cut}^{(2)}$ (eq. \ref{eq:cut}),
until the outflow expands to a radius,
\begin{equation}
 R_m=3\times10^{13}L_{52}\Gamma_{2.5}^{-3}\rm cm
\end{equation}
(note $R_m>R_\gamma$). At $R>R_m$, the cutoff energy turns to increase with $R$,
$\varepsilon_{\rm cut}\simeq\varepsilon_0 R/R_m$. In both cases of $\Gamma>\Gamma_c$
and $\Gamma<\Gamma_c$ the cutoff energy at $R>\max[R_\gamma,\,R_m]$ (i.e,
$\varepsilon_{\rm cut}>\varepsilon_0$) follows the same expression,
\begin{equation}\label{eq:cut evolution}
  \varepsilon_{\rm cut}^{\rm ob}\simeq50\frac{t_{\rm var,-2}^{\rm ob}\Gamma_{2.5}^6}{L_{52}}\pfrac
  R{10^2R_\gamma}(1+z)^{-2}\rm GeV.
\end{equation}
The evolution of $\varepsilon_{\rm cut}$ versus $R$ for fixed $L$ is plotted in fig
\ref{fig:cutvsR}.

Comparing $\varepsilon_{\rm IC}$ and $\varepsilon_{\rm cut}$ it
can be found that typically $\varepsilon_{\rm IC}>\varepsilon_{\rm
cut}$ at small radii, $R\la30R_\gamma$. In this case the bulk IC
radiation is absorbed in the source, leading to electromagnetic
cascades, and the photons escape until their energies decay to
$\varepsilon_{\rm cut}$. Therefore the bulk high energy radiation
is just re-emitted at $\varepsilon_{\rm cut}$. On the other hand,
$\varepsilon_{\rm IC}<\varepsilon_{\rm cut}$ at large radii
$R\ga30R_\gamma$, where only the high energy emission below
$\varepsilon_{\rm cut}$ appears. The emission above
$\varepsilon_{\rm cut}$ is truncated, and undergoes
electromagnetic cascades, but does not affect much the apparent
spectrum since the photon spectrum decreases rapidly with energy.

Finally consider the emission flux. It is straightforward to show
that the cooling time of the electrons is short compared to the
dynamical time during the late residual collision phase, up to
radii $R\sim10^3R_\gamma$. We therefore assume that electrons
radiate away all their energy. When $\varepsilon_{\rm
IC}>\varepsilon_{\rm cut}$ at small radii, the total electron
energy appears at $\varepsilon_{\rm cut}\simeq\varepsilon_0$ if
$R_\gamma<R<R_m$ or $\varepsilon_{\rm cut}\propto R$ if
$R>\max[R_\gamma,R_m]$. The observed (time-integrated) IC spectrum
at energy $\varepsilon>\varepsilon_0$ would be $\nu F_\nu\propto
E_{\rm fluc}|_{\nu_{\rm cut}=\nu}\propto R^{-2/3}|_{\nu_{\rm
cut}=\nu}\propto\nu^{-2/3}$ (Here $\nu=\varepsilon/h$).

When the outflow expands to large radii where $\varepsilon_{\rm
IC}<\varepsilon_{\rm cut}$, we need to consider the electrons
accelerated to Lorentz factors larger than the characteristic
Lorentz factor $\gamma_i$. Shock acceleration is expected to
generate a power-law energy distribution of electrons
$dn_e/d\gamma_e\propto\gamma_e^{-p}$ at $\gamma_e>\gamma_i$ with
$p\simeq2$ \citep{BE87}. This flat-electron energy distribution,
$\gamma_e^2dn_e/d\gamma_e\propto\gamma_e^0$, generates equal
amounts of IC energy in logarithmic photon energy intervals, $\nu
F_\nu\propto \nu^0$ for $\nu>\nu_{\rm IC}$. So $\nu F_\nu(\nu_{\rm
cut})\simeq\nu F_\nu(\nu_{\rm IC})\propto E_{\rm fluc}$. The
emission at low energy would be covered by earlier emission, while
only the emission at high energy end, i.e. around the cutoff,
shows up and interests us. The observed (time-integrated) spectrum
would be similar to the $\varepsilon_{\rm IC}>\varepsilon_{\rm
cut}$ case, i.e.,  $\nu F_\nu\propto \nu^{-2/3}$.

\begin{figure}
\includegraphics[width=\columnwidth]{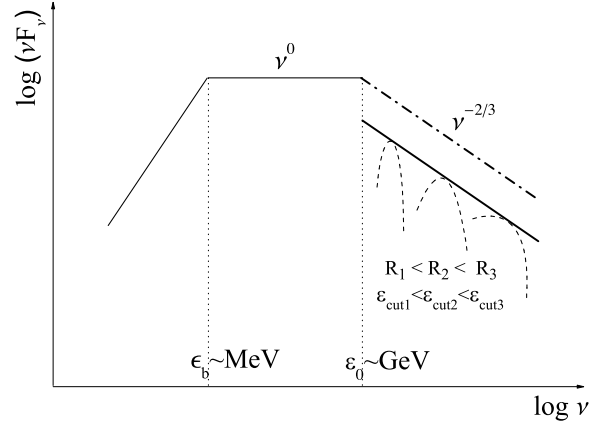}
\caption{Schematic plot of the predicted $\nu F_\nu$ spectrum of
prompt high energy emission from a GRB. The {\em thin solid} line
shows the observed MeV \gmr emission, a broken power law with a
break energy at $\eps_b\sim$~MeV, above which the spectrum goes as
$\nu F_\nu\propto\nu^0$. The {\em dot} lines mark the break energy
$\eps_b$ and the previously thought pair-production spectral
cutoff, $\varepsilon_0\sim$~GeV (eq. [\ref{eq:cut}]). The residual
collisions at large radii contribute beyond $\varepsilon_0$.
Summing up all emission components (the {\em dashed} lines) from
different radii and times lead to a spectral slope $\nu
F_\nu\propto\nu^{-q}$ ($q$ is the index of the random energy
evolution, $E_{\rm fluc}\propto R^{-q}$, and $0<q\leq2/3$). In the
simplified case, $q=2/3$ (see the text). Here the {\em thick
solid} line corresponds to the outflow satisfying $R_\gamma<R_m$
(or $\varepsilon_0\simeq\Gamma m_ec^2$), while the {\em thick
dashed dot} line corresponds to $R_\gamma>R_m$
($\varepsilon_0\ga\Gamma m_ec^2$). The $\nu F_\nu$ values in these
two are different by a factor of $(R_m/R_\gamma)^{2/3}$. The
spectral slope holds up to TeV range, but not higher (see the
discussion section). } \label{fig:spectrum}
\end{figure}

Thus we expect the observed (time-integrated) prompt spectrum, above the spectral
cutoff energy in the prompt \gmr emitting phase, $\varepsilon_0$ (eq. \ref{eq:cut}),
to be
\begin{equation}\label{eq:spec}
   \nu F_\nu\propto\nu^{-2/3}, ~~h\nu>\varepsilon_0~~~(\rm simplified ~~case).
\end{equation}
This fluence spectrum is resulted from summing up all emission
components from different radii and times. A schematic plot of the
prompt GRB spectrum is shown in fig. \ref{fig:spectrum}. Note, the
spectral slope $\nu^{-2/3}$ here is derived from the simple case,
which has been confirmed by recent numerical calculation by
\cite{Aoi09}. If in general $E_{\rm fluc}\propto R^{-q}$ we would
expect the slope to be $\nu F_\nu\propto \nu^{-q}$.

Below $\varepsilon_0$ is the observed prompt MeV \gmr spectrum,
i.e., typically $\nu F_\nu\propto\nu$ below $\eps_b$ and $\nu
F_\nu\propto\nu^0$ between $\eps_b$ and $\varepsilon_0$. Note, the
transition of the emission from primary to residual collisions at
$\varepsilon_0$ is smooth if $R_\gamma>R_m$, as shown by the
dashed dot line. The transition for the case of $R_\gamma<R_m$ is
discontinuous as shown by the thick solid line. The power law
described by eq. (\ref{eq:spec}) starts with a flux lower than the
primary emission at $\varepsilon_0$ by a factor of
$(R_m/R_\gamma)^{2/3}$.

Some comments should be made here. The spectral form described in
eq. (\ref{eq:spec}) holds only on average, especially for the high
energy range. In individual GRB events the flux may differ
significantly, because, for a small number of shells (and
collisions) at large radii, large variations in the late residual
collisions should be expected.

It should also be noticed that we have assumed the initial
variance $\sigma_{\Gamma,0}<\Gamma$, whereas initial condition
with $\sigma_{\Gamma,0}>\Gamma$ may lead to more efficient \gmr
production \citep[e.g.][]{Blbrdv00} around $R_\gamma$, in which
case the ratio between fluxes of primary and residual emission at
$\varepsilon_0$ should be larger by a factor of a few\footnote{It
is not expected that the initial variance of Lorentz factors is
far exceeding the mean, $\sigma_{\Gamma,0}\gg\Gamma$.}, leading to
more abrupt transition between the primary and residual emission.

One may worry about that the IC emission may be reduced as the
seed photons are not isotropic in the shock frame of residual
collisions. However, this kind of geometry effects do not play an
important role even if the photons are completely collinear
\citep{Wang06}. As usually assumed by many authors, suppose that
the electrons accelerated in residual collision shocks are
isotropic in the rest frame of the outflow, since the tangled
magnetic fields in the shock might sufficiently isotropize
electrons. Thus an electron is changing its angle $\theta'$ with
respect to the photon beam and cooling fast. The IC power of an
electron averaged over its cooling time is not different from
interacting with isotropic photons. As long as the jet effect is
not important to prompt GRB emission ($\theta_{\rm
jet}\gg1/\Gamma$), we can furthermore regard the GRB explosion as
isotropic, and hence the observer at different angles will observe
the same IC emission due to spherical symmetry. Consider both
cases of isotropic and anisotropic scatterings, the radiated
energy can be assumed to be the same because it is determined by
the total electron energy if electrons radiate all their energy
rapidly. The outside observers would observe the same fluence in
both cases, otherwise one can simply ask where the electron energy
have gone, given the same total electron energy. Thus, the IC
fluence is not reduced by this geometry effect if electrons are
isotropic distributed in the rest frame and radiate all their
energy within a dynamic time. Nevertheless, this effect changes
the angular distribution of IC emission. In the rest frame the IC
power becomes $P_{\rm IC}'\propto(1-\cos\theta')$, although not
much different from isotropic distribution. Correspondingly, in
the lab frame the ``image'' of the IC emission is different from
the isotropic case, i.e. the anglular dependence of the brightness
is different, but the angular-integrated fluence is the same.

\subsubsection{Time delay}
At energy $\varepsilon<\varepsilon_0$, the emission is mainly
contributed by primary collisions at small radius, and arrives at
detectors simultaneously with MeV emission. However, for higher
energy $\varepsilon>\varepsilon_0$, the emission is produced at
relatively large radii, and should have a time delay relative to
the primary MeV emission. Since the spectral cutoff energy, where
the high energy photons emerge, increases as the outflow expands,
$\varepsilon_{\rm cut}\propto R$ for $\varepsilon_{\rm
cut}>\varepsilon_0$ (eq. \ref{eq:cut evolution}), the time delay
increases with observed photon energy. For energy
$\varepsilon=\varepsilon_{\rm cut}(R)>\varepsilon_0$, the typical
radius where photons emerge is
$R\simeq[\varepsilon/\varepsilon_0]\max[R_\gamma,\,R_m]=[\varepsilon/\varepsilon_{\rm
cut}^{(1)}]R_\gamma$, therefore the related time delay relative to
MeV \gmr emission, $\tau_{\rm delay}\simeq
(R-R_\gamma)/2\Gamma^2c$, is
\begin{equation}\label{eq:delay}
  \tau_{\rm delay}^{\rm ob}\simeq\frac{\varepsilon}{\varepsilon_{\rm
cut}^{(1)}}t_{\rm
var}=2.2L_{52}\Gamma_{2.5}^{-6}\frac{\varepsilon^{\rm ob}}{\rm
10^2~GeV}(1+z)^2\rm
  s~~(for ~\varepsilon^{\rm ob}>\varepsilon_0^{\rm ob}).
\end{equation}
Thus the emission at $\varepsilon>\varepsilon_0$ is delayed later
as $\varepsilon$ increases. A comment is made here that the target
photons for scattering may be beamed with respect to electrons,
which changes the angular distribution of IC emission, i.e. the
maximum IC power may come from a certain angle other than
$\theta=0$. This leads to an additional time delay. However, as
implied by eq. (\ref{eq:max_angle}), the angle where the maximum
power is emitted is smaller than $\pi/2$ in the comoving frame and
hence $\theta<1/\Gamma$. The produced time delay is smaller than
$\tau_{\rm delay}$, $R\theta^2/2c<R/2\Gamma^2c\sim \tau_{\rm
delay}$. Thus we neglect this additional time delay.

Eq.\,(\ref{eq:delay}) implies that the detection of time delay
$\tau_{\rm delay}$ at $\varepsilon$ helps to determine the Lorentz
factor,
\begin{equation}\label{eq:Gamma_delay}
  \Gamma=167L_{52}^{1/6}\pfrac{\varepsilon^{\rm ob}}{1\rm\,GeV}^{1/6}\pfrac{\tau_{\rm delay}^{\rm ob}|_{\varepsilon^{\rm ob}}}{1\,\rm
  s}^{-1/6}(1+z)^{1/3}.
\end{equation}
This determination by time delay should be consistent with that by
detection of the spectral transition $\varepsilon_0$ between
primary and residual emission, eq.\,(\ref{eq:Gamma_cut}).

\section{Application: Fermi-LAT GRB 080916c}
As the Fermi observational data showed up after the first version
of this paper was posted on the
archive,\footnote{http://arxiv.org/abs/0810.2932} we now apply the
model to the observations.

\cite{Fermi916c} reports the measurements of the bright GRB
080916c by Fermi GBM and LAT. The redshift of this burst is
$z=4.35$ \citep{afterglow916c}, which implies, with flucence (10
keV$-$10 GeV) $\approx2.4\times10^{-4}\rm ~erg\,cm^{-2}$, the
largest reported isotropic \gmr energy release, $E_{\rm
iso}\simeq9\times10^{54}$erg. The observed GRB duration is
$T\approx50$~s, so the bolometric isotropic-equivalent luminosity
is $L_{\rm bol}=E_{\rm iso}(1+z)/T\approx10^{54}\rm erg\,s^{-1}$.
As the observed peak energy is $\eps_b/(1+z)\sim1$~MeV and the
high energy slope is $\beta\approx-2$, the MeV luminosity, defined
as the luminosity at $\la2\eps_b$, is $L\approx L_{\rm
bol}/\ln(10\,\rm GeV/1\,MeV)=10^{53}\rm erg\,s^{-1}$. The LAT
detected 145 photons at $>100$~MeV, within which 14 are beyond 1
GeV, during the first 100 s after the trigger. The brightness may
have enough statistics for spectral and temporal analysis of the
high energy properties.

There are several interesting properties in the high energy
emission of this GRB.

\paragraph{Time delay}
{\em The multi-band light curves unambiguously show that the bulk
of the emission of the second light-curve peak is moving toward
later times as the energy increases \citep[see time bin b in Fig 1
and its inset panels in ][]{Fermi916c}, and the time delay of
$100$-MeV emission is about 1 s relative to MeV emission, much
larger than the MeV variability timescale, $\la100$~ms
\citep{afterglow916c}.}\footnote{Note, the time-delay issue here
is different from what is called "delayed onset" by other authors.
We concern indeed the delayed peaking time of high-energy
emission, related to the delayed arrival of the bulk of
high-energy emission.} First of all, these are qualitatively
consistent with our prediction that the higher energy photons can
only arise when the plasma expands to larger size in later time
where the $\gamma\gamma$ optical depth reduces to below unity, and
that the size of high energy emission can be much larger than MeV
emission.

Let us consider the data quantitatively, and constrain the model parameters. The LAT
>100 MeV detection consists of 145 photons which mainly come up in a single light-curve peak,
therefore we have enough statistics for the time analysis of >100
MeV emission. It is obviously seen that the >100 MeV light curve
peak has a time delay $\tau_{\rm delay}^{\rm ob}=\tau_{\rm
delay}(1+z)\sim1$~s relative to that of 250~keV-5~MeV (The script
"ob" denotes quantities measured in the observer frame, with
redshift effect taken into account). This implies
$\varepsilon_0^{\rm ob}=\varepsilon_0/(1+z)<100$~MeV. Substituting
the observed values of $L=10^{53}\rm erg\,s^{-1}$, $\tau_{\rm
delay}\simeq1/5.35$~s and $\varepsilon=100\times5.35=535$~MeV,
with redshift $z=4.35$ taken, into eq.\,(\ref{eq:delay}), we
obtain the bulk Lorentz factor of GRB 080916c outflow,
\begin{equation}\label{eq:916c_LF}
  \Gamma\simeq290\pfrac{L}{10^{53}\rm erg\,s^{-1}}^{1/6}\pfrac{\tau_{\rm delay}^{\rm
  ob}|_{\rm 100\,MeV}}{1~\rm s}^{-1/6}.
\end{equation}
This result is similar to those determined in other GRBs,
$\Gamma\approx100-400$, through observations of the rising of
optical afterglows \citep{LFopt1,LFopt2,LFopt4,LFopt3}, and the
thermal components in the prompt emission \citep{peer07}.

The determination of $\Gamma$ can be double checked by the
location of $\varepsilon_0$. Using the result of
eq.\,(\ref{eq:916c_LF}), we obtain $$\varepsilon_{\rm cut}^{(1)\rm
obs}\simeq10(t_{\rm var}^{\rm ob}/10^2\rm ms)(\tau_{\rm
delay}^{\rm ob}|_{\rm 100\,MeV}/1\,\rm s)^{-1}\rm MeV$$ and
$$\varepsilon_{\rm cut}^{(2)\rm ob}\simeq40(L/10^{53}\rm
erg\,s^{-1})^{1/6}(\tau_{\rm delay}^{\rm ob}|_{\rm
100\,MeV}/1\,\rm s)^{-1/6}\rm MeV.$$ The observed MeV variability
timescale is $t_{\rm var}^{\rm ob}\la$100 ms based on the INTEGRAL
observation \citep{afterglow916c}. Thus $\varepsilon_0^{\rm
ob}\sim40$~MeV, consistent with requirement $\varepsilon_0^{\rm
ob}<100$~MeV. In addition, the broad light curve peak in the
no-energy-selection band of LAT is consistent with, or a little
delayed from, that of the GBM (260 keV-5 MeV) light curve, and is
ahead of the >100 MeV peak. Thus $\varepsilon_0^{\rm ob}$ should
be located in the LAT energy window (no selection) and below 100
MeV, consistent with the result $\varepsilon_0^{\rm
ob}\sim40$~MeV.

By our model the >1~GeV emission should be even 10 times longer delayed than the >100
MeV one, i.e., $\tau_{\rm delay}^{\rm ob}|_{1\,\rm GeV}\sim10$~s. The much fewer
photons above 1 GeV prevent us from analyzing the temporal properties with high
confidence. However the LAT >1 GeV light curve does agree with a longer delay by
$\sim10$~s.

It should be noticed that other authors also constrain the bulk
Lorentz factor of this GRB and obtain much larger lower limit
\citep{afterglow916c,Fermi916c}. Essentially, the difference is
due to different models considered; they consider the GeV emission
produced in the same time and place as the MeV emission, whereas
in our model the GeV emission comes from delayed residual
collisions at large radii, therefore our model looses the
constraint on $\Gamma$. In addition, we consider that the cutoff
energy should not locate below $2^{1/2}\Gamma m_ec^2$, which is
ignored in \cite{afterglow916c} and \cite{Fermi916c}.

\paragraph{Lack of the first LAT light-curve peak}
The low energy GBM light curve shows two peaks, however the LAT
observations only show one peak related to the second GBM peak and
there is a paucity of emission in the first $\sim4$~s after the
trigger. Note, some people call this as a "delayed onset" of high
energy emission. In principle, one of the explanations could be
that there is a spectral cutoff at $\sim10$~MeV for the prompt
emission from primary collisions in the first 4 s, and the
residual-collision emission at >10 MeV is $\ga4$~s delayed, longer
than the second peak. If so, the properties of the ejecta emitting
the first GBM peak are different from the later ejecta, which also
suggests that there might be long-timescale, a-few-second ($\gg
t_{\rm var}\sim1$~ms), variabilities in the outflow of this GRB.

\paragraph{Time-integrated spectrum}
The joint GBM-LAT spectrum of GRB 080916c can be fit by Band
function \citep{Band93}, with peak energy around 1~MeV,
$\alpha\approx-1.0$ and $\beta\approx-2.2$, except for the first 4
s \citep{Fermi916c}. Because the time intervals used to construct
the spectra are much longer than the MeV variability time, the
resulted spectra are all time-integrated ones. Since the
synchrotron self-Compton (SSC) model for GRBs, where the MeV peak
is from the IC scattering by soft photon emitting electrons,
predicts a bright GeV-TeV component due to the second order
up-scattering \citep{Piran08}, no evidence for high energy bump up
to 10 GeV in observations does not favor SSC but synchrotron model
\citep{Wang09}. In addition, the narrow $\nu F_\nu$ spectral peak
of GRB spectra favor more synchrotron emission mechanism over SSC,
since SSC usually has a much broader spectral bump
\citep{Baring04}.

In the framework of synchrotron internal shocks, our residual
collision model predicts a slight spectral softening at high
energies. However, due to the small detected GeV-photon number the
Poisson scatter of low statistic still allows a slight softening
at tens MeV to fit the data. Furthermore, if there are large
timescale, $\gg1$~ms, variabilities in the outflow so that the
residual emission spectral slope $q$ is larger than $2/3$ (see
\S\ref{section:simulation}), the high-energy spectrum is less
steepened and is closer to a single power law. Finally, the slope
of eq.(\ref{eq:spec}) holds on average, while the later residual
collisions occur between smaller number of shells, thus there
might be fluctuation from this average slope.

\paragraph{Highest energy emission}
The highest energy photon is detected with 13.2 GeV only 17 s
after the GRB trigger. With the redshift $z=4.35$ this suggests
GRB 080916c produces radiation up to 71 GeV in the source frame.
Moreover, LAT detects 145 photons with $>100$~MeV, within which 14
with $>1$~GeV and especially only one with $>10$~GeV, consistent
with a power law spectrum with photon index $\beta\approx-2$ up to
$\sim10$~GeV scale. There might be emission extending to higher
energy, say, beyond tens of GeV, from GRB 080916c following the
same slope, but the detection rate is less than one, i.e., no
photon would be detected at this energy. Thus the observations
actually suggest that the high energy spectral cutoff (or steep
drop), if there exists, is more likely to be far above the energy
of the only observed highest energy photon,
$\varepsilon_{\max}^{\rm ob}=13.2\times g$~GeV with $g\gg1$. If
the high energy emission beyond 13.2 GeV is produced by internal
shocks, we argue here that it may not be produced by primary
internal collisions that emit MeV \gmrs but produced in other
regions, e.g., by residual collisions.

As said above no high energy spectral component in GRB 080916c
does not favor SSC model but synchrotron model. Now calculate the
maximum synchrotron photon energy. If $B$ is the magnetic field
strength in the internal shock, the Larmor time of an electron
with Lorentz factor $\gamma$ is $t_L'=\gamma m_ec/eB$. The typical
particle acceleration time can be scaled by Larmor time as
$t_a'=ft_L'$ \citep[e.g.][]{Hillas84} where $f$ is a correction
factor accounting for the uncertainty of shock acceleration. It
might be that $f\ga$ a few \citep[e.g.][]{Lemoine06}. In the same
time the electron suffers synchrotron cooling in a typical
timescale $t_c'=3m_ec/4\sigma_T\gamma(B^2/8\pi)$ (We neglect the
IC cooling as the IC scattering usually occurs in deep
Klein-Nishina regime for the most energetic electrons). The
competition between acceleration and cooling results in a maximum
Lorentz factor of accelerated electrons, $\gamma_{\max} =(6\pi
e/f\sigma_TB)^{1/2}$. The relevant synchrotron photon frequency is
a constant,\footnote{This upper bound for synchrotron energy is
robust for any acceleration mechanisms involving electromagnetic
processes, because the acceleration limit with $f=1$ is robust not
only to Fermi shock accelerations but also to any particle
accelerations through electromagnetic processes. Therefore this
bound might be valid not only to internal shock models but also to
electromagnetic-dominated models.}
$$\nu_{\max}'=0.3\gamma_{\max}^2eB/2\pi
m_ec=0.9e^2/f\sigma_Tm_ec.$$ The coefficient accounts for the fact
that the synchrotron power per unit frequency of an electron peaks
at the frequency a fraction 0.3 of the common characteristic
frequency. The maximum synchrotron photon energy
$\varepsilon_{\max}^{\rm ob}=h\nu_{\max}'\Gamma/(1+z)$ is,
therefore,
\begin{equation}
  \varepsilon_{\max}^{\rm ob}=15\Gamma_{2.5}f^{-1}(1+z)^{-1}\rm
  ~GeV.
\end{equation}

Comparing the predicted maximum synchrotron energy with that
implied by the observation, $\varepsilon_{\max}^{\rm
ob}=13.2g$~GeV, we have a lower bound,
\begin{equation}
  \Gamma=1.5\times10^4f_{0.5}g_{0.5},
\end{equation}
where the conservative values, $f=10^{0.5}f_{0.5}$ and
$g=10^{0.5}g_{0.5}$, have been taken. This bulk Lorentz factor is
too large for fireball-shock model, because it faces several
problems. First, the large $\Gamma$ leads to (primary) internal
shock radius larger than the deceleration radius of GRB outflow
\citep[e.g.][]{Lazzati99}. The deceleration radius is
$R_d\approx(E_k/4nm_pc^2\Gamma^2)^{1/3}$. If $R_\gamma<R_d$, an
upper limit is obtained,
\begin{equation}
  \Gamma<7\times10^3(E_{k,55}/n_0)^{1/8}t_{\rm var,-2}^{-3/8},
\end{equation}
where $E_k=10^{55}E_{k,55}$erg and $n=1n_0\rm cm^{-3}$ are the outflow kinetic energy
and medium density, respectively. Second, the large $\Gamma$ raises problem of low
energy conversion efficiency due to slow cooling of accelerated electrons
\citep[e.g.,][]{Derishev01}. If the synchrotron cooling time of electrons with typical
postshock Lorentz factor $\gamma_m\sim m_p/m_e\sim10^3\gamma_3$, is required to be
smaller than the dynamical time of the outflow, $t_c'(\gamma_m)<t_d'\simeq
R_\gamma/\Gamma c$, we have
\begin{equation}
  \Gamma<5\times10^3\gamma_3^{1/5}(L_{\rm bol}/10^{54}{\rm erg\,s^{-1}})^{1/5}t_{\rm
  var,-2}^{-1/5}.
\end{equation}
In this calculation we have assumed that the postshock magnetic
field is limited by observed emission, $B^2/8\pi\leq
U_\gamma=L_{\rm bol}/4\pi R_\gamma^2\Gamma^2c$. Third, in
synchrotron internal shock models the large $\Gamma$ leads to
large collision radius $R_\gamma\approx2\Gamma^2ct_{\rm var}$,
where the magnetic field $B$ is too small to give rise high energy
synchrotron photon energy. Using
$\eps_b\approx\Gamma\hbar\gamma_m^2eB/m_ec$ and the limit
$B^2/8\pi< U_\gamma$, the restriction to obtain synchrotron
emission peaking at MeV range is
\begin{equation}
  \Gamma<0.4\times10^3\gamma_3(L_{\rm bol}/10^{54}{\rm erg\,s^{-1}})^{1/4}t_{\rm
  var,-2}^{-1/2}(\eps_b/\rm 1\,MeV)^{-1/2}.
\end{equation}
Finally, the thermal pressure of the initial fireball is not
expected to accelerate the loaded baryons to very large Lorentz
factor with most energy kept as the kinetic energy of baryons. The
final Lorentz factor is limited to be
\begin{equation}
  \Gamma<3\times10^3(L_0/10^{54}{\rm erg\,s^{-1}})^{1/4}r_{0,7}^{-1/4},
\end{equation}
where $L_0$ is the rate at which the central source emits energy,
and $r_0=10^7r_{0,7}$~cm is the source size \citep[see,
e.g][]{Waxman rev}.

The above upper limits to $\Gamma$ imply that it may be impossible
that $\Gamma\gg10^3$. This appear not to match the value suggested
by the highest energy band observation,
$\Gamma=1.5\times10^4f_{0.5}g_{0.5}$, unless $f\approx1$ and
$g\approx1$ are satisfied at the same time: the shock acceleration
must operate at the Bohm limit; furthermore the observed highest
energy photon happens to be at the maximum synchrotron energy.
Thus the observations imply there is emission much higher than
13.2 GeV, which cannot be originated from synchrotron emission in
the primary internal shocks. Actually, this high energy emission
can be produced by IC emission in residual collision shocks, as
discussed in present study. So the observation of highest energy
emission supports the residual collision model.

In conclusions: (1) the time delay of high energy emission and the
spectral feature of highest energy emission in GRB 080916c might
have provide evidences for the residual collision model; (2) its
spectrum is not inconsistent with the residual emission; (3) the
time delay of $>100$ MeV emission constrains the bulk Lorentz
factor to be $\Gamma\sim300$, a typical value usually taken. It
appears to be an applicable method to determine $\Gamma$ of GRB
outflows by measuring the time delays of LAT light curves. If
internal shocks also work in short GRBs, we expect similar
delayed, prompt high energy emission in short GRBs. We also
caution more careful spectral analysis to find the transition
between primary and residual emission.

\section{Discussion}
We have considered the high energy emission in the prompt GRB
spectrum, which is dominated by the IC emission from the residual
collisions. Instead of a exponential spectral cutoff, a steeper,
compared to the prompt MeV emission, power-law slope $\nu
F_\nu\propto \nu^{-q}$ is expected beyond the previously thought
cutoff, $\varepsilon_0$ (eq. [\ref{eq:cut}]). Here $q$ is
corresponding to the dynamical evolution of the random energy in
the outflow, $E_{\rm fluc}\propto R^{-q}$. In the simplified case
(see \S 3.1), which is consistent with optical flash observations
(LW08), we take $q=2/3$, while $0< q\leq 2/3$ in general. The
extended emission makes it complicated to detect the "cutoff
energy" in the goal to constrain the GRB emission region.

Indeed, EGRET had detected prompt high energy photons past GeV in
several brightest BATSE GRBs occurring in its field of view (e.g.,
GRB 930131, \cite{930131}; GRB 940217, \cite{Hurley94}), which
suggest that the other faint GRBs may produce prompt GeV photons
as well \citep{Dingus95}. There is also no sign of cutoff in the
spectra \citep{Dingus95}, which, if there is, should be far
exceeding $\sim1$~GeV. These EGRET results are consistent with the
predicted extension of prompt emission beyond GeV. However, the
cutoff is not ruled out. Given the sensitive dependence of the
cutoff on $\Gamma$ (eq.\ref{eq:cut}), a slight variation of
$\Gamma$ in individual GRB events may lead to much higher cutoff
energy, $\gg1$~GeV, explaining the prompt GeV emission in
EGRET-detected GRBs. Two properties may help to discriminate our
residual-collision emission model from a very high energy cutoff
model. The first is the steepening turnover in the spectrum. For
typical Lorentz factors, $\Gamma\approx10^2-10^3$, the expected
spectral turnover is $\rm \sim100~MeV-1~TeV$, well located in the
windows of Fermi and AGILE. The second is the time delay of high
energy emission. One may expect systematic time delay of the high
energy photons in the residual emission model, while no delay is
expected in the very high energy cutoff model. However, the task
is not easy given that for a typical event with fluence
$\sim10^{-6}\rm erg~cm^{-2}$ the observed GeV photon number is
only a few. In order to have enough statistics for the spectral
and temporal analyses, very bright events are needed, or one may
integrate many events to obtain an average burst.

We have discussed that the recent Fermi detected bright GRB
080916c might have presented a good sample. Observations do not
show a simple spectral cutoff, but a spectral tail up to 70 GeV in
GRB frame. More than one hundred of photons detected above 100 MeV
makes it obviously showing a time delay about 1 sec, which can be
explained by the residual IC emission and results in the
determination of a typical Lorentz factor value, $\Gamma\sim300$.
The features of GRB 080916c support the residual-collision
emission model as opposed to the very-high-energy-cutoff model.

The high energy emission would not extend to very high energy.
There are several effects that lead to a drop in TeV range. First,
when the plasma expands to very large radius,
$R\ga2\Gamma^2cT\sim10^{17}\Gamma_{2.5}^2(T/10{\rm~s})$~cm ($T$ is
the MeV \gmr duration), there would be no overlapping between the
plasma and the MeV \gmrs, and hence no scattering is expected. At
$R\simeq2\Gamma^2cT$ the cutoff energy increases to
$\varepsilon_{\rm cut}\simeq1(T/10{\rm~s})t_{\rm var,-2}^{-1}$~TeV
(from eq.\ref{eq:cut evolution}). A lack of $\ga1$~TeV photons
would be expected for GRBs with duration $T\la10t_{\rm var,-2}$~s,
although it should not be an exponential cutoff. Second, in
interaction with $\sim1$~MeV photons, the Klein-Nishina limit
becomes important for electrons with $\gamma_e\ga\Gamma$, giving
rise to IC photons up to $\varepsilon_{\rm
IC}\simeq0.1\Gamma_{2.5}^2$TeV, beyond which the spectrum
gradually turns below the low energy slope. Finally the cosmic
infrared background would absorb the $>0.1$~TeV photons that
arrive from GRBs at redshift $z\ga0.5$. Except for low redshift
events, the observed prompt GRB emission may not extend far beyond
TeV range.

It should be commented here that in the framework of the
synchrotron internal shock model \citep{Waxman rev} we do not
expect a bright high energy component, say $\ga 1$~GeV, in the
prompt emission, other than the synchrotron self-Compton model
\citep[e.g.][]{Piran08}, therefore the residual high energy
emission will be dominant. A detection of high energy component in
the prompt emission will be an evidence against the synchrotron
model for GRBs, and vice verse. The recent Fermi observations of
several GRBs do not support a high energy component in GRB
spectrum, since the $>100$~MeV fluences are all less than those in
MeV range \citep{Fermi916c}.

Our residual collision model is not expected to produce much
longer delayed high energy emission which is not apparently
overlapped with the primary MeV \gmrs in times. There are more and
more observations showing that GRBs produce delayed $>100$~MeV
emission even after the prompt \gmrs end, lasting tens or even
$\sim10^4$ seconds \citep{Hurley94,dermer03,Giuliani08}. This may
require some long-lasting central engine activities or external
productions \citep[e.g.][and references therein]{Wang06}.

\acknowledgments
This work was partly supported by the National
Natural Science Foundation of China through grants 10473010 and
10843007.

\appendix
\section{Dynamics with multi-timescale
variabilities}\label{section:simulation}

In order to show the effect of multi-timescale variabilities on
the dynamics of the outflow, we carry simulations for both cases
of single- and multi-timescale variabilities for comparison. We
consider a series of individual material shells $i=1,2...,N$, with
total shell number $N=3000$, released in a duration of $T=3$s, so
that the interval between two nearby shells is $1$~ms. The shells
have equal masses but different energies, with the bulk Lorentz
factor of each shell following
\begin{equation}
  \log\Gamma_i=2+\xi_i\log9+\phi_i,
\end{equation}
where $\xi_i$ is a random number between zero and unity and
$\phi_i$ can be taken as the following forms,
\begin{equation}
  \phi_i=\left\{\begin{array}{ll}
    0 & \rm (Single);\\
    \sum_k  \sin(\frac{2\pi Ti}{P_kN})\log A_k & \rm
    (Multi).
  \end{array}
  \right.\label{eq:multi-scale}
\end{equation}
For single-timescale case, we should take $\phi_i=0$ (case S),
then the outflow has only one variability timescale of $10^{-3}$s.
The Lorentz factors randomly and uniformly distributed in
logarithmical scale between 100 and 900. For multi-timescale
effect we consider 3 extra timescales besides the smallest
timescale: $P_k=10^{-2},10^{-1},1$~s for $k=1,2,3$, with relevant
$A_k$ values being $A_k=2,1.5,1.2$ (case M1) or $A_k=1.2,1.5,2$
(case M2) for $k=1,2,3$. $A_k$ decreases with $P_k$ in case M1 but
increase in case M2, which means case M2 has larger fluctuations
at larger timescales. Larger timescale fluctuations tend to
produce strong collisions at larger radii so that larger fraction
of energy is dissipated at larger radii. We further consider that
in each two-shell collision, $1/3$ of the generated internal
energy is emitted by radiation, because only the energy of shocked
electrons is assumed to be emitted rapidly and the electron
equipartition parameter is $\epsilon_e=1/3$. The two shells are
considered to separate again after collision and share equally the
remained internal energy (in the center-of-momentum frame of the
two shells).
\begin{figure}[h]
\includegraphics[width=\columnwidth]{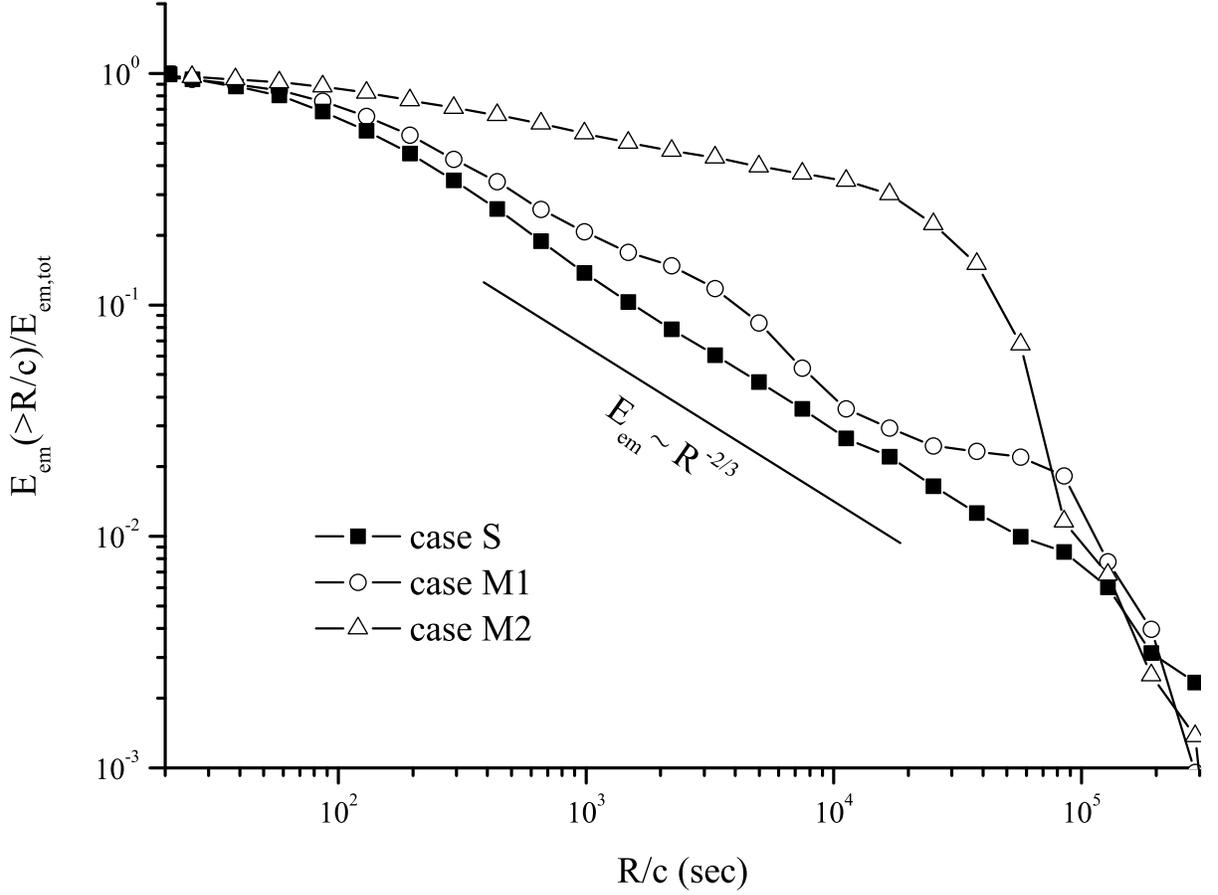}
\caption{The fraction of emission energy as function of radius in
three simulations. The line marked with {\em squares}: case S,
with only the 1-ms variability; {\em circles}. The other two are
multi-timescale cases (M1 and M2), with three more scales of
$P_k=10^{-2}, 10^{-1}$ and 1 s, for $k=1, 2,$ and 3, respectively.
The line marked with {\em circles}: case M1, with $A_k=2, 1.5,$
and 1.2 for $k=1,2,$ and 3 (eq.[\ref{eq:multi-scale}]); {\em
triangles}: case M2, with $A_k=1.2, 1.5,$ and 2 for $k=1, 2,$ and
3 (eq.[\ref{eq:multi-scale}]). The straight line show the $-2/3$
slop to guide eyes. See the relevant text for more
details.}\label{fig:simulation}
\end{figure}

In Fig \ref{fig:simulation} we show the fraction of emission
energy $E_{\rm em}(>R)/E_{\rm em,tot}$, that is emitted beyond
radius $R$. We see that case S show a slope close to the
analytical resolution $\propto R^{-2/3}$ for single-timescale case
by LW08. However, the multi timescales lead to flatter slopes,
implying more fraction of emission energy tends to be emitted at
larger radii. Case M2 has even flatter slope than case M1, since
case M2 has relatively larger fraction of energy that is
dissipated at larger radii. The steep drop at the end in both
cases M1 and M2 means no more strong collisions later on. This is
because there is a largest timescale of 1~s in our simulations. If
there is still variabilities with timescale larger than 1 s then
the slope will continue to even larger radii and show even later
drop at the end. In summary, the simulations demonstrate that
multi-timescale variabilities lead to a flatter slope $q<2/3$, and
the $q=2/3$ slope is only for single-timescale case.

\section{Anisotropic IC emission}
Let us discuss at which energy the IC emission is emitted in the
observer frame, taking into account the fact that the seed photons
are beamed. Consider the extremely anisotropic case, where the
photons are collinear in the comoving frame of the outflow. In
this frame the electrons, as argued, is reasonably assumed to be
isotropically distributed. For simplicity, we consider
mono-energetic photons, since the photon number rapidly decreases
with energy. Thus the IC power per unity solid angle in the
comoving frame is angular dependent,
\begin{equation}
  \frac{\d P'}{\d \Omega'}\propto(1-\mu')^2,
\end{equation}
 where $\mu'=\cos\theta'$
with $\theta'$ the direction with respect to photon beam, and we
have taken the velocity of the electron to be $\beta_e'\approx1$.
Hereinafter prime denotes quantities in the comoving frame of the
outflow, while non-prime denotes observer frame. Using the Lorentz
transformation, $\mu'=(\mu-\beta_\Gamma)/(1-\beta_\Gamma\mu)$,
where $\beta_\Gamma=(1-1/\Gamma^2)^{1/2}$, we have
\begin{equation}
  1-\mu'=(1+\beta_\Gamma)\frac{1-\mu}{1-\beta_\Gamma\mu},
\end{equation}
then the angular distribution of IC power in observer frame is
\begin{equation}
  \frac{\d P}{\d \Omega}=\frac 1{\Gamma^4(1-\beta_\Gamma\mu)^3}\frac{\d P'}{\d \Omega'}
  \propto\frac{(1-\mu)^2}{(1-\beta_\Gamma\mu)^5}.
\end{equation}
This is not like the simple cone-like distribution of the
isotropic-photon case. The maximum power per solid angle is
emitted at angle with
\begin{equation}
  \mu_{\max}=\frac{5\beta_\Gamma-2}{3\beta_\Gamma}.
\end{equation}
The corresponding angle in the comoving frame is given by
\begin{equation}\label{eq:max_angle}
  1-\mu_{\max}'=\frac{2(1+\beta_\Gamma)}{5\beta_\Gamma}\approx\frac45.
\end{equation}
The scattered photon energy in the comoving frame is given by
$\varepsilon_{\rm IC}'\approx\gamma_e^2\eps'(1-\mu')$, with
$\gamma_e$ the electron Lorentz factor and $\eps'$ the photon
energy both in the comoving frame. The photon energy (in observer
frame) emitted at angel $\mu_{\max}$, where the IC power is
maximum, is then
\begin{equation}
  \varepsilon_{\rm IC}(\mu_{\max})=\frac{\varepsilon_{\rm
IC}'(\mu_{\max})}{\Gamma(1-\beta_\Gamma\mu_{\max})}=\frac{3\gamma_e^2\eps'}{5\Gamma(1-\beta_\Gamma)}
  \approx\frac35\gamma_e^2\Gamma\eps'(1+\beta_\Gamma)\approx\frac35\gamma_e^2\eps,
\end{equation}
where in the last equality $\eps=(1+\beta_\Gamma)\Gamma\eps'$ is
taken for collinear photons. This is the observer-frame photon
energy around which the IC emission is mainly emitted. We see that
the anisotropic correction factor $\lambda$ in eq.
(\ref{eq:ICph_energy}) is only order unity, implying that taking
$\lambda\sim1$ is a good approximation even for highly beamed seed
photons.

\end{document}